\documentclass[journal]{IEEEtran}

\usepackage{amssymb,amsmath,amsthm,enumitem}
        \usepackage{setspace}

\usepackage{graphicx}
\usepackage{caption}
\usepackage{subcaption}
\usepackage[labelformat=parens,labelsep=quad,skip=3pt]{caption}

\ifCLASSINFOpdf
  
\else
 
\fi

\usepackage{amsmath}
\usepackage{amssymb}  
\usepackage{amsfonts}
\usepackage{mathrsfs}
\usepackage{mathptmx} 
\usepackage{mathtools}

\usepackage{xcolor}

\usepackage{algorithm}
\usepackage{algpseudocode}
\usepackage[algo2e]{algorithm2e} 

\usepackage{multirow}
\usepackage{subcaption}
\usepackage{csquotes}

\begin{document}
%
\title{Data-Driven Safe Gain-Scheduling Control }

%


\author{Amir~Modares,~\IEEEmembership{Student~Member,~IEEE,}
        Naser~Sadati,~\IEEEmembership{Senior~Member,~IEEE,}
       and~Hamidreza~Modares,~\IEEEmembership{Senior~Member,~IEEE}  
\thanks{M. Shell was with the Department
of Electrical and Computer Engineering, Georgia Institute of Technology, Atlanta,
GA, 30332 USA e-mail: (see http://www.michaelshell.org/contact.html).}
\thanks{J. Doe and J. Doe are with Anonymous University.}
 \thanks{This work was supported by Ford Motor Company-Michigan State University Alliance.}
\thanks{
A. Modares and N. Sadati are with the Department of Electrical Engineering, Sharif University of Technology, Tehran, Iran. 
(e-mails: amir.modares.81@gmail.com, sadati@sharif.edu).}
}


%
%

\maketitle

\begin{abstract}
Data-based safe gain-scheduling controllers are presented for discrete-time linear parameter-varying systems (LPV) with polytopic models. First, $\lambda$-contractivity conditions are provided under which safety and stability of the LPV systems are unified through Minkowski functions of the safe sets. Then, a data-based representation of the closed-loop LPV system is provided, which requires less restrictive data richness conditions than identifying the system dynamics. This sample-efficient closed-loop data-based representation is leveraged to design data-driven gain-scheduling controllers that guarantee $\lambda$-contractivity, and, thus, invariance of the safe sets.  It is also shown that the problem of designing a data-driven gain-scheduling controller for a polyhedral (ellipsoidal) safe set amounts to a linear program (a semi-definite program). The motivation behind direct learning of a safe controller is that identifying an LPV system requires at least $nsm+ns+m$ samples to satisfy the persistence of excitation (PE) condition, where $n$ and $m$ are dimensions of the system's state and input, respectively, and $s$ is the number of scheduling variables. It is shown in this paper, however, that directly learning a safe controller and bypassing the system identification can be achieved without satisfying the PE condition. This data-richness reduction is of vital importance, especially for LPV systems that are open-loop unstable, and collecting rich samples to satisfy the PE condition can jeopardize their safety. A simulation example is provided to show the effectiveness of the presented approach. 

\end{abstract}

 \vspace{-10pt}

%
\IEEEpeerreviewmaketitle

\section{Introduction}
%
%
%
%
\IEEEPARstart{S}{atisfaction} of safety constraints is a fundamental requirement for control systems that must be deployed on safety-critical systems, such as autonomous vehicles and robots. 
Design of safe controllers using barrier certificates \cite{CB1}--\cite{CB11} and reachability analysis \cite{Reach1}--\cite{Reach5} has been widely and successfully considered. These methods, however, mostly rely on a high-fidelity model of the system under control. To account for model uncertainties, robust safe control methods \cite{Robust1}, \cite{Robust2} design a controller for the worst-case uncertainty realization. Worst-case-based control design, nevertheless, can result in overly-conservative control solutions and even infeasibility. On the other hand, adaptive safe control methods \cite{Adapt1}, \cite{Adapt2} are designed to compensate for uncertainties and avoid conservatism. These methods, however, are based on the availability of an adaptive control Barrier function (aCBF), which is challenging to find for nonlinear systems. 
 Safe reinforcement learning (RL) algorithms have also been presented in \cite{SafeRL1}--\cite{SafeRL7} to learn constrained optimal control solutions for systems with uncertain dynamics. Nevertheless, a model-based safety certifier is used in safe RL methods to intervene with the RL actions whenever they are not safe.
 
Performance and conservativeness of barrier-based safety certifiers and controllers highly depend on the accuracy of the identified system model. However, as shown in \cite{datainf1}, \cite{datainf2}, conditions imposed on the data richness for identifying a linear system model are generally more restrictive than conditions imposed on data richness for directly learning a controller that satisfies a system property (e.g., stability). Therefore, to avoid data-hungry learning developments, it is desired to directly design safe controllers using measured data along the system trajectories. Moreover, if stability is also of a concern in these methods, a control Lyapunov function (CLF) based constraint is also typically imposed besides a control barrier function. When there is a conflict between safety and stability, however, the CLF is relaxed, which can result in the convergence of the closed-loop trajectories to an undesired equilibrium point on the boundary of the unsafe set \cite{conflict}. To avoid this conflict, the concept of contractive sets \cite{SetB} can be leveraged for linear systems with convex safe sets to unify the safe and stable control design. This idea is leveraged in \cite{Data2}-\cite{Data4} to directly design data-driven safe controllers for linear time-invariant systems. The data-based safe control design is also considered in \cite{Data1} using only measured data collected from open-loop system trajectories. Existing results for direct data-driven safe control design are limited to linear time-invariant systems and impose restrictive requirements on data richness. 

Designing data-driven safe and stable controllers for general nonlinear systems with general safety constraints is a daunting challenge. However, many nonlinear systems such as aerospace systems \cite{LPV1}, \cite{LPV2} and a variety of robotic systems \cite{LPV3}, \cite{LPV4} can be expressed by linear-parameter varying (LPV) systems with a set of gain-scheduling parameters that are not known in advance, but can be measured or estimated during operation of the system. Safe and stable gain-scheduling control strategies can then be unified for LPV systems under convex and compact constraint sets  \cite{SetB}.

This paper presents data-based safe gain-scheduling controllers for LPV systems with both ellipsoidal and polytopic safe sets. Our approach is inspired by \cite{Data2}, which is presented for linear time-invariant systems with ploytopic safe sets, and extends its results to nonlinear LPV systems with both closed and convex polyhedral and ellipsoidal safe sets. Moreover, it is shown here that the data requirement conditions for directly learning a safe controller are actually weaker than the standard persistence of excitation (PE) condition. That is, even when data samples do not satisfy the PE condition and thus identifying the LPV dynamics is not possible, the presented approach can directly learn a safe controller if the non-PE data satisfy a relaxed condition. When exploration to generate rich data is risky, this direct data-driven approach with a lower sample complexity will be highly advantageous to model-based approaches that rely on system identification.

To design direct data-driven safe gain-scheduling controllers, first,  $\lambda$-contractivity conditions are provided for LPV systems with known dynamics under which the safety and stability of the LPV systems are guaranteed for both ellipsoidal and polyhedral safe sets through their related Minkowski functions. Then, to obviate the requirement of knowing the system dynamics, a data-based representation of the closed-loop system is provided, making evident how this parametrization is naturally related to the $\lambda$-contractive sets. The set invariance and stability of the LPV systems are then guaranteed through Minkowski functions. It is shown that the problem of designing controllers to enforce a given polyhedral set and an ellipsoidal set to be $\lambda$-contractive in the presented data-based framework amounts to a linear program and a semi-definite program, respectively. A motivation example shows that while a safe controller can be learned using available data measurements, the data is not rich enough to identify the LPV model.
Finally, a simulation example is provided to show the effectiveness of the presented approach. \vspace{3pt}

\noindent \textbf{Notations}: Throughout the paper, $\otimes$ denotes the Kronecker product and $ \odot $ denotes the Khatri--Rao product, which is a column-wise Kronecker product of two matrices that have an equal number of columns. ${I}$ denotes the
identity matrix with appropriate dimension, $\textbf{0}_n$ denotes the $n \times n$ zero matrix and
$\bar 1$ denotes the vector of all ones of appropriate dimension. If $A$ and $B$ are matrices (or vectors) of the same dimensions, then $A (\le,\ge) B$ implies a componentwise inequality, i.e.,  ${A_{ij}} (\le,\ge) {B_{ij}}$ for all $i$ and $j$, where $A_{ij}$ is the element of the $i$-th and $j$-th column of $A$. In the space of symmetric matrice, $Q \preceq  0$ denotes that  $Q$ is negative semi deﬁnite. Moreover, $A^{\dag}$ is the right inverse of the matrix $A$. Given a polyhedron $A$, vert ($A$) denotes the set of its vertices. Given a set $\cal{S}$ and a scalar $ \mu  \ge 0$, the set $\mu \cal{S}$
is defined as   $\mu \cal{S}$:= 
$\{ \mu x:x \in \cal{S}\} $.  \vspace{3pt}

\noindent \textbf{Definition 1.} \cite{SetB} A convex and compact set that includes the origin as its interior point is called a C-set. \vspace{3pt}

\noindent \textbf{Definition 2.} The set $\varepsilon (P,1)$ is an ellipsoidal C-set and is represented by \vspace{-6pt}
\begin{align} \label{elip}
\varepsilon (P,1) = \{ x \in {R^n}:\sqrt {{x^T}Px}  \le 1\}, 
\end{align}
where $P$ is a positive definite matrix.\vspace{3pt} 

\noindent \textbf{Definition 3.} A polyhedral C-set 
${\cal {S}} (F,\bar 1)$  is represented by
\begin{align} \label{poly}
{\cal {S}} (F, \bar 1) = \{ x \in {R^n}:Fx \le \bar 1\}  = \{ x \in {R^n}:{F^i}x \le 1,i = 1,...,q\},	
\end{align} 
where $F \in {R^{q \times n}}$ is a matrix with rows ${F^i},i = 1,...,q$. 

\section{Problem Formulation} 
This section formulates problems of safe control design for polytopic LPV systems with  both polyhedral and ellipsoidal safe sets.  

Consider the discrete polytopic LPV system given by \cite{MLPV1}, \cite{MLPV2} \vspace{-6pt}
\begin{align} \label{LPVsys}
x(t + 1) = A(w(t))x(t) + Bu(t),
\end{align}	
where $x(t) \in {X}$ is the system’s state and $u(t) \in {\cal{U}}$ is the control input with $X$ and ${\cal{U}}$ as constrained sets (e.g., ellipsoidal or polyhedral) containing the origin in their interiors. Moreover,
$B \in {R^{n \times m}}$ is the input dynamic and is assumed fixed. The parameter-varying matrix 
$A(w(t))$  is known to lie in the following polytope 	\vspace{-9pt}
 \begin{align} \label{AWs}
 A(w(t)) = \sum\limits_{i = 1}^s {{A_i}} {w_i}(t),
 \end{align}	
 where 
${A_i} \in {R^{n \times n}}, \, i = 1,...,s$ are vertices of the polytope and
$w \in \Omega $  is a scheduling parameter vector. While the scheduling parameter can be measured online (e.g., velocity of an aircraft), its future values are not known and are supposed to belong to the following polytope.
\begin{align}
\Omega  = \{ w \in {R^s},{w_i} \ge 0,\sum\limits_{i = 1}^s {{w_i}}  = 1\}.
\label{omega set}
\end{align}

The gain-scheduling controller is typically considered as $u(t) = K(w)x(t)$, where \vspace{-6pt}
\begin{align} \label{KG}
 K(w) = \sum\limits_{i = 1}^s {{K_i}} {w_i}. 
\end{align} 

\noindent \textbf{Assumption 1.}  The number of operating modes, i.e., $s$, is known. 
This can be prior knowledge or the knowledge obtained through clustering of the data samples collected from the system’s trajectories, as performed in \cite{Sadati}.\vspace{6pt}



\noindent\textbf{Problem 1.} Given a polyhedral C-set ${\cal {S}} (F, \bar 1)$, find the gain-scheduling controller $u(t) = K(w)x(t)$, with $K(w)$ defined in \eqref{KG}, such that it guarantees the following: \vspace{6pt}

\noindent1)	The set ${\cal {S}} (F, \bar 1)$  remains invariant.

\noindent2)	The origin is asymptotically stable.\vspace{6pt}

\noindent\textbf{Problem 2.} Given an ellipsoidal C-set 
$\varepsilon (P,1)$, find the gain-scheduling controller $u(t) = K(w)x(t)$, with $K(w)$ defined in \eqref{KG}, such that it guarantees the following: \vspace{6pt}

\noindent1)	The set $\varepsilon (P,1)$  remains invariant.

\noindent2)	The origin is asymptotically stable.\vspace{6pt}

In Problems 1 and 2, the first property guarantees system safety and the second property guarantees its stability. In this paper, gain-scheduling controllers are designed to solve Problems 1 and 2 based on only the trajectories of data measurements collected from system's inputs, states, and scheduling parameters. \vspace{3pt}

\noindent\textbf{Remark 1.} For both polyhedral and ellipsoidal C-sets, the safety and stability properties of LPV systems can be embedded in the notion of $\lambda$-contractivity, defined next. This will significantly simplify designing controllers that are both safe and stable and can avoid converging to an undesired equilibrium solution of the closed-loop system that can arise due to the conflict between safety and stability in barrier-certificate based approaches \cite{conflict}.  \vspace{6pt}

\noindent\textbf{Definition 4:} (Minkowski function) Given a C–set $\cal{S}$, its Minkowski function is
\begin{align}  
\Psi _{\cal{S}}(x) = \inf \{ \alpha  \ge 0:x \in \alpha \cal{S} \}.
\end{align}

\noindent\textbf{Definition 5:} (Contractive set) Fix $\lambda  \in [0,1)$. The C-set  $\cal{S}$ is $\lambda$-contractive for the system 
$x(t + 1) = f(x(t),t)$ if and only if $x(0) \in \cal{S}$ implies that $x(t) \in \lambda \,\, \cal{S}$, $\forall t \ge 0$. 

For a $\lambda$-contractive set $\cal{S}$, the following condition holds \cite{SetB}
\begin{align} \label{Minkowski}
\Psi _{\cal{S}}(f(x,t)) \le \lambda,  \,\, {\forall x \in \cal{S}}, 
\end{align}
where $\Psi _{\cal{S}}(x)$ is the Minkowski function of $\cal{S}$. \vspace{3pt}

The following results show that the  Minkowski function is actually a (local/global) shared control Lyapunov function that guarantees both stability and safety of the LPV systems with (constrained/unconstrained) inputs. \vspace{3pt}

\noindent\textbf{Theorem 1.}  \cite{SetB} Let $u\in {R^m}$ be unbounded and the C–set $\cal{S}$ (e.g., polyhedral or ellipsoidal) be $\lambda$-contractive for the closed-loop system \eqref{LPVsys}. Then, the system is both safe and stable. Moreover, let $\Psi_{\cal{S}}(x)$  be the Minkowski function for the set $\cal{S}$. Then, $\Psi _{\cal{S}}(x)$ is a global control Lyapunov function. \vspace{3pt}


\noindent\textbf{Corollary 1.} \cite{SetB} Let $u \in {\cal{U}}$  be bounded and the C–set $\cal{S}$(e.g. polyhedral or ellipsoidal) be $\lambda$-contractive for the closed-loop system \eqref{LPVsys}. Then, the system is both safe and stable. Moreover, let $\Psi _{\cal{S}}(x)$  be the  Minkowski function for the C–set $\cal{S}$. Then,  $\Psi _{\cal{S}}(x)$  is a control Lyapunov function inside the set  $\cal{S}$. \vspace{-6pt}
\section{Model-Based Controller Design for Solving Problems 1 and 2}  \vspace{-2pt}
It was shown in Theorem 1 that to solve Problem 1, it is sufficient to design a controller that guarantees that the set ${\cal{S}}(F, \bar 1)$ is $\lambda$-contractive. The next results provide conditions under which the $\lambda$-contractiveness is guaranteed for both Polyhedral C-sets and ellipsoidal C-sets. 

Before proceeding, the following notations are defined and used throughout the paper for the system \eqref{LPVsys}, \eqref{AWs}.
\begin{align} \label{K_1}
  \nonumber {K_{1,s}} = \left[ {{K_1},{K_2},...,{K_s}} \right],  \\
  {A_{1,s}} = \left[ {{A_1},{A_2},...,{A_s}} \right], 
\end{align}
where ${\rm{ }}{K_i} \in {R^{m \times n}}$. This gives
\begin{align} \label{ABK}
  {A_{1,s}} + B{K_{1,s}} = \left[ {{A_1} + B{K_1},...,{A_s} + B{K_s}} \right].
\end{align}  \vspace{-23pt}
\subsection{ Model-Based Solving of Problem 1 for Polyhedral Sets} 
\noindent We present the next result on $\lambda$-contractivity for polytopic models under polyhedral C-set constraints on their states.\vspace{3pt}
\noindent\textbf{Theorem 2.} Consider the LPV system \eqref{LPVsys} with  $s$ vertices and a polyhedral C-set  ${\cal{S}}(F, \bar 1)$ of the form \eqref{poly}. Let
$u(t) = K(w) x(t)$ with $K(w)$ defined in \eqref{KG}. Then, the C-set ${\cal{S}}(F, \bar 1)$   is  $\lambda$-contractive for closed-loop system \eqref{LPVsys} if and only if there exists a non-negative matrix 
${P_{1,s}} = \left[ {{P^1},{P^2},...,{P^s}} \right]$ such that
\begin{align} \label{modelpoly}
 \nonumber
&{\rm{ }}{P_{1,s}}({\delta _j} \otimes \bar 1) \le \lambda \bar 1, \,\, j = 1,...,s \\
&{P_{1,s}}({I} \otimes F) = F({A_{1,s}} + B{K_{1,s}}),
\end{align}
where 
 ${K_{1,s}}$ and ${A_{1,s}} + B{K_{1,s}}$ are defined in \eqref{K_1} and \eqref{ABK}, respectively, and ${\delta _j}\in {R^s}$ is a vector with all elements zero except its $j\mbox{-}th$ element, which is one.

\noindent\textit{Proof.} The closed-loop polytopic LPV system \eqref{LPVsys} is  $\lambda$-contractive if and only if it is $\lambda$-contractive in its $s$ vertices \cite{SetB}. Therefore, the closed-loop polytopic LPV system \eqref{LPVsys} with  $u(t) = K(w)x(t)$ is $\lambda$-contractive if and only if there exist $s$  non-negative matrices  
${P^i} \ge 0$ and ${K_i}$  satisfying 
\begin{align}
 {P^i}F = F({A_i} + B{K_i}), \,\,\, {\rm{  }}{P^i}\bar 1 \le \lambda \bar 1, \,\,\, {\rm{  }}i = 1,...,s.
\end{align}

\noindent Compounding the above inequalities and equalities, respectively, yields
\begin{align}
& \nonumber  \left[ {{P^1},{P^2},...,{P^s}} \right]({\delta _j} \otimes \bar 1) \le \lambda \bar 1, \,\,\, {\rm{}}j=1,...,s\\ 
&   \left[ {{P^1},{P^2},...,{P^s}} \right]\left[ \begin{array}{l}
{I} \otimes F
\end{array} \right] = F\left[ {{A_1} + B{K_1},...,{A_s} + B{K_s}} \right],
\end{align}
which yields \eqref{modelpoly} all together using ${A_{1,s}} + B{K_{1,s}}$ in \eqref{ABK}. This completes the proof. \hfill   $\square$ \vspace{-6pt}



\subsection{Model-Based Solving of Problem 2 for Ellipsoidal Sets}

\noindent 
We present the next result on  $\lambda$-contractivity for polytopic models under ellipsoidal constraints.\vspace{3pt}

\noindent\textbf{Theorem 3.} Consider the LPV system \eqref{LPVsys} and an ellipsoidal C-set
$\varepsilon (P,1)$ of the form $\eqref{elip}$.
Let $u(t) = K(w) x(t)$ with $K(w)$ defined in \eqref{KG}. Then, the C-set $\varepsilon (P,1)$   is  $\lambda$-contractive for closed-loop system \eqref{LPVsys} if and only if 
\begin{align} \label{modelelip}
 {D_i}^T{({A_{1,s}} + B{K_{1,s}})^T}P({A_{1,s}} + B{K_{1,s}}){D_i} - {\lambda ^2}P \preceq 0, \,\,\, i = 1,...,s,
\end{align}

\noindent where 
${D_i} \in {R^{(n \times s) \times n}},i = 1,...,s$ and
\begin{align}
\begin{array}{l}
{D_i}^T = \left[ {\textbf{0}_n, \cdots ,\underbrace { {{I_n}}}_{i \mbox{-} th \,\,{\rm{ n}} \times {\rm{n \,\, matrix}}},\textbf{0}_n, \cdots,\textbf{0}_n } \right].
\end{array} \label{D matrix}
\end{align}

\noindent\textit{Proof.} The LPV system \eqref{LPVsys} with  
$u = K(w) x$  is  $\lambda$-contractive with respect to C-set 
$\varepsilon (P,1)$  if and only if  \cite{SetB}
\begin{align}
{({A_i} + B{K_i})^T}P({A_i} + B{K_i}) - {\lambda ^2}P \preceq 0{\rm{   }},\,\,\, i = 1,...,s.   
\end{align} \vspace{-6pt}

\noindent Defining $D_i$ matrices as \eqref{D matrix} and using \eqref{ABK}, this condition reduces to \eqref{modelelip}. \hfill   $\square$ \vspace{-5pt}
\section{Data-Based Representation of LPV systems} \vspace{-3pt}

\noindent This section provides a data-based representation of LPV systems. 

Solving Problems 1 and 2 using Theorems 2 and 3 requires the complete knowledge of all the dynamics 
${A_1},{A_2},...,{A_s}$  and $B$,  which are not known in advance. This paper presents a data-based solution to Problems 1 and 2 to obviate the need for this knowledge. That is, the set invariance and stability of LPV systems are imposed without the knowledge of the system matrices and only by relying on a finite number of data samples collected from the inputs, states and scheduling parameters. The data samples are collected by applying a sequence
${u_d}(0),...,{u_d}(T \mbox{-} 1)$ of inputs and measuring the corresponding values 
${x_d}(0),...,{x_d}(T)$ for a measured sequence of  
${w_d}(0),...,{w_d}(T \mbox{-} 1)$,  where the subscript $d$  emphasizes that these are data. A single data set that spans over a large range of operating conditions (a rich set of scheduling variables) and a rich set of state and input data are now organized as follows. 
\begin{align}
 & {U_0} = \left[ {{u_d}(0),...,{u_d}(T \mbox{-} 1)} \right] \label{datU}\\
 & {X_0} = \left[ {{x_d}(0),...,{x_d}(T \mbox{-} 1)} \right]  \label{DatX0}\\
 &  {X_1} = \left[ {{x_d}(1),...,{x_d}(T)} \right] \label{datX1} \\
 &  {W_0} = \left[ {{w_d}(0),...,{w_d}(T \mbox{-} 1)} \right] \label{datW1} \\
 &  {X_W} = {W_0} \odot {X_0}. \label{datW}
\end{align}
This richness condition for the collected dataset is investigated next. A rich data set will allow us to design data-based controllers that capture the dependency structure of the matrices of the LPV state-space model on the scheduling variables without requiring an explicit model or declaration of dependencies. \\

\noindent\textbf{Remark 2.} A promising data-based safe control design approach is presented in \cite{Data2} for linear time-invariant systems. However, it is not investigated in \cite{Data2} how the direct learning of a safe controller can reduce the sample complexity (i.e., the number of samples required to learn) compared to learning a system model first and then designing a model-based safe controller. That is, their developments are based on the assumption that the collected data satisfy the PE requirement. Satisfying the PE requirement for the LPV systems amounts to having
the data matrix  
\begin{align} \label{rank}
\left[ \begin{array}{l}
{U_0}\\
{X_W}
\end{array} \right], 
\end{align}
with full row rank. That is, the number of samples in \eqref{datU}--\eqref{datW} must satisfy $T \ge nms+ns+m$. As shown in the next theorem, inspired by \cite{datainf}, this condition provides sufficient for uniquely identifying the LPV system. Once the system is identified, the results of Theorems 2 and 3 can be used to design a model-based controller. However, as shown later, one can learn directly a data-based safe controller using less restrictive data informative conditions. Therefore, it is more desirable to directly learn a safe controller. \vspace{2.5pt}

\noindent \textbf{Theorem 4.} The LPV system \eqref{LPVsys} can be uniquely identified if the matrix \eqref{rank} is full row rank. Moreover, under this condition, it has the following equivalent data-based representation
\begin{align} \label{dataLPVO}
 x(t + 1) = {X_1}{\left[ \begin{array}{l}
{U_0}\\
{X_W}
\end{array} \right]^\dag }\left[ \begin{array}{l}
{\rm{     }}u{{(t)}}\\
w{\rm{(}}t{\rm{)}} \otimes x{{(t)}}
\end{array} \right].
\end{align}
\noindent \textit{Proof.} Based on \eqref{LPVsys}, the data collected in \eqref{datU}--\eqref{datW} satisfy
\begin{align}
&\nonumber {X_1} = {A_{1,s}}\,\, [{w_d}(0) \otimes {x_d}(0),...,{w_d}(T \mbox{-} 1) \otimes {x_d}(T \mbox{-} 1)] +\\  
&B \,\, [{u_d}(0),...,{u_d}(T \mbox{-} 1)] = [B{\rm{  }}\,\,\,\,{{\rm{A}}_{1,s}}]\left[ \begin{array}{l}
{U_0}\\
{X_W}
\end{array} \right]{\rm{ }}, \label{eq1}
\end{align}
where $A_{1,s}$ is defined in \eqref{K_1}. There exists a right inverse $[V_1 \,\,\, V_2]$ such that \vspace{-4pt}
\begin{align} \label{inv}
\left[ \begin{array}{l}
{U_0}\\
{X_W}
\end{array} \right] [V_1 \,\,\, V_2]=
I,
\end{align}
if and only if the matrix \eqref{rank} is full row rank. In this case, multiplying both sides of \eqref{eq1} by $[V_1 \,\,\, V_2]$, one can uniquely find ${A_{1,s}}$ and $B$ as ${A_{1,s}}=X_1 V_1$ and $B=X_1 V_2$. We now show that \eqref{dataLPVO} holds. Based on \eqref{LPVsys}, one has \vspace{-1pt}
\begin{align} \label{eq5}
x(t + 1) = \left[ {B{\rm{  }}\,\,\,\,{{\rm{A}}_{1,s}}} \right]\left[ \begin{array}{l}
u(t)\\
w(t) \otimes x(t)
\end{array} \right].
\end{align}
On the other hand, 
\begin{align} \label{g}
\left[ \begin{array}{l}
u\\
w \otimes x
\end{array} \right] = \left[ \begin{array}{l}
{U_0}\\
{X_W}
\end{array} \right]g,
\end{align}
admits a solution $g$ given by
\begin{align} \label{eq4}
g = [V_1 \,\,\, V_2] \left[ \begin{array}{l}
u\\
w \otimes x
\end{array} \right] + \Big(I - [V_1 \,\,\, V_2] \left[ \begin{array}{l}
{U_0}\\
{X_W}
\end{array} \right]\Big)d,\,\,\,\,
\end{align}
for any ${d} \in {R^T}$, where 
$\Big(I - [V_1 \,\,\, V_2] \left[ \begin{array}{l}
{U_0}\\
{X_W}
\end{array} \right]\Big)$  is the orthogonal projector onto the kernel of 
$\left[ \begin{array}{l}
{U_0}\\
{X_W}
\end{array} \right]$. Using \eqref{g} in \eqref{eq5}, one has \vspace{-6pt}
\begin{align} \label{eq55}
x(t + 1) =  \left[ {B{\rm{  }}\,\,\,\,{{\rm{A}}_{1,s}}} \right]\left[ \begin{array}{l}
{U_0}\\
{X_W}
\end{array} \right]g(t).
\end{align}
Using \eqref{eq1} and \eqref{eq4} this becomes
\begin{align}
x(t + 1) = {X_1}\bigg([V_1 \,\,\, V_2] \left[ \begin{array}{l}
u(t)\\
w(t) \otimes x(t)
\end{array} \right] + \Big(I - [V_1 \,\,\, V_2]\left[ \begin{array}{l}
{U_0}\\
{X_W}
\end{array} \right]\Big)d\bigg),
\end{align}
where \vspace{-6pt}
\begin{align}
    {X_1}\bigg(I - [V_1 \,\,\, V_2]\left[ \begin{array}{l}
{U_0}\\
{X_W}
\end{array} \right]\bigg)=0,
\end{align}
which completes the proof. \hfill   $\square$ \vspace{3pt}

\noindent \textbf{Remark 3}. Theorem 4 provides a data-based representation that predicts the system's state for any given input. However, in the safe control design, one only needs the data-based closed-loop representation of the system for a state-feedback controller that must be designed to assure safety. Therefore, instead of requiring both $B$ and $A_{1,s}$ to be implicitly known, only $A_{1,s}+BK_{1,s}$ must be implicitly known through data for a specific data-dependent $K_{1,s}$. Therefore, the rank condition requirement in Theorem 4 can be relaxed for the data-based closed-loop representation with a data-dependent $K_{1,s}$. The data richness requirement for the safe control design under which the gain $K_{1,s}$ can be obtained from the closed-loop representation is presented next. \vspace{3pt}

\noindent \textbf{Assumption 2} The matrix $X_W$ has full row rank. \vspace{3pt}

Note that since $X_W \in \mathbb{R}^{ns \times (T-1)}$, satisfying the full row rank condition of Assumption 2 requires $T-1 \ge ns$, or, equivalently, $T \ge ns+1$. \vspace{3pt} 

\noindent\textbf{Theorem 5.}  Let Assumption 2 hold. Then, the closed-loop system \eqref{LPVsys} with the gain-scheduling controller $u(t) = K(w)x(t)$, where $K(w)$ is defined in \eqref{KG}, has the following representation \vspace{-6pt}
\begin{align} \label{LPVdata1}
  x(t + 1) = {X_1}{G_{{K_{1,s}}}}(w(t) \otimes x(t)),
\end{align}
or equivalently 
\begin{align} \label{closedloop}
   {A_{1,s}} + B{K_{1,s}} =  {X_1}{G_{{K_{1,s}}}},
\end{align}
\noindent where 
${G_{{K_{1,s}}}} \in {R^{T \times (s \times n)}}$ satisfies
\begin{align} \label{G}
\left[ \begin{array}{l}
{K_{1,s}}\\
{I}
\end{array} \right] = \left[ \begin{array}{l}
{U_0}\\
{X_W}
\end{array} \right]{G_{{K_{1,s}}}},
\end{align}
where $K_{1,s}$ is defined in \eqref{K_1}.

\noindent\textit{Proof.}  Since $X_W$ has full row rank, there exits a right inverse $G_{K_{1,s}}$ such that \vspace{-6pt}
\begin{align} \label{GW}  
 X_W \, G_{K_{1,s}}=I.
\end{align}
By applying the input sequence \eqref{datU} and the scheduling sequence \eqref{datW1} to the LPV system \eqref{LPVsys}, one has
\begin{align} \label{data lpv}
{X_1} = {A_{1,s}}{X_W} + B{U_0}.
\end{align}
Multiplying $G_{K_{1,s}}$ to both sides of \eqref{data lpv} from right gives
\begin{align}\label{data form}
{X_1}G_{K_{1,s}} = {A_{1,s}} + B{U_0}G_{K_{1,s}},
\end{align}
On the other hand, from \eqref{G}, the control gain is ${K_{1,s}} = {U_0}G_{K_{1,s}}$. Therefore, \eqref{data form} becomes
\begin{align}\label{data form1}
{X_1}G_{K_{1,s}} = {A_{1,s}} + B \, K_{1,s},
\end{align}
which is equivalent to \eqref{closedloop}. Moreover, the system \eqref{LPVsys} with the gain-scheduling control law 
$u = \sum\limits_{i = 1}^s {{K_i}} {w_i}x$  transforms to
\begin{align} \label{cl}
 x(t + 1) = \sum\limits_{i = 1}^s {({A_i}}  + B{K_i}){w_i}x(t) = ({A_{1,s}} + B{K_{1,s}})(w(t) \otimes x(t)).
\end{align}
Using \eqref{data form1} in \eqref{cl} gives \eqref{LPVdata1}. This completes the proof. \vspace{5pt}



\section{Data-Driven Safe Gain-Scheduling Control for LPV Systems}

\noindent Theorem 5 showed that the closed-loop gain-scheduling system is parameterized through data via \eqref{LPVdata1}, \eqref{G}. Since the matrix ${G_{{K_{1,s}}}}$ in the closed-loop representation of Theorem 5 is not unique, the next results will treat it as a decision variable to design data-based safe controllers. \vspace{-9pt}

\subsection{Data-Based Safe Gain Scheduling for Polyhedral Sets} \vspace{-2pt}
In this subsection, the data-based closed-loop representation provided in Theorem 5 is leveraged to directly design safe gain-scheduling controllers for a given polyhedral set. \vspace{6pt}

\noindent\textbf{Theorem 6.} Consider the data collected in \eqref{datU}--\eqref{datW}. Let Assumptions 1 and 2 hold. Let
$u \in {R^m}$. Then, Problem 1 is solved if there exist decision variables 
${G_{{K_{1,s}}}}$ and
${P_{1,s}} \ge 0$ such that \vspace{-8pt}
\begin{subequations}
\begin{align} \label{databasedpolyF}
&P_{1,s} ({\delta _j} \otimes \bar 1) \le \lambda \bar 1, \,\, j = 1,...,s \\ \label{databasedpolyF1}
&P_{1,s} (I \otimes F) = F{X_1}{G_{{K_{1,s}}}} \\
&{X_W}{G_{{K_{1,s}}}} = I \label{databasedpolyF2}.
\end{align}
\end{subequations}
Moreover, 
${K_{1,s}} = {U_0}{G_{{K_{1,s}}}}$, and thus the control gains that solve the problem are obtained as 
${K_i} = {U_0}{G_{{K_{1,s}}}}{D_i}$.  \vspace{1pt}

\noindent\textit{Proof.} It was shown in Theorem 2 that to solve Problem 1, the gain matrix  
${K_{1,s}}$ must satisfy \eqref{modelpoly}. Therefore, the proof is completed if one shows that satisfying \eqref{databasedpolyF}--\eqref{databasedpolyF2} implies satisfying \eqref{modelpoly} with 
${K_{1,s}} = {U_0}{G_{{K_{1,s}}}}$. The inequalities in both equations are identical. Using  \eqref{databasedpolyF2} and ${K_{1,s}}={U_0}{G_{{K_{1,s}}}}$, one has \eqref{G}, which has a solution based on Theorem 5 and under Assumpion 2. Comparing the second equation of \eqref{modelpoly} and \eqref{databasedpolyF1}, the proof is completed if we show that the term ${A_{1,s}} + B{K_{1,s}}$ in \eqref{modelpoly} is equal to ${X_1}{G_{{K_{1,s}}}}$. This 
 is shown in Theorem 5 under \eqref{G}, and thus the proof is completed. \hfill   $\square$  \vspace{6pt}
 
 \noindent\textbf{Remark 5.} By Theorem 1, if there exist decision variables 
${P_{1,s}}$ and 
${G_{{K_{1,s}}}}$ that satisfy (27), then the closed-loop LPV system is globally asymptotically stable and Minkowski function of the polyhedral set ($\max {}_{i = 1,...,s}({F^i}x)$) is a global Lyapunov function. \vspace{3pt}

Theorem 6 relies on the closed-loop representation provided in Theorem 5, which requires Assumption 2 to be satisfied on data. The following example shows that while the data is not rich enough for system identification, it can be used to directly design a safe controller. \vspace{3pt}  

 \noindent \textbf{Example}
Consider a polytopic LPV system in the form of \eqref{dataLPVO} with
\begin{align}
{A_1} = \left[ \begin{array}{l}
.4 \,\,\,\,\,\,\,\,\,\,\,\, {\rm{     0}}\\
{\rm{0   \,\,\,\,\,\,\,   - }}{\rm{.1}}
\end{array} \right],\,\,\, {A_2} = \left[ \begin{array}{l}
{\rm{ - }}{\rm{.3 \,\,\,\,\,\,\,    0}}\\
{\rm{1    \,\,\,\,\,\,\,\,\,\,\,\,\,   }}{\rm{.1}}
\end{array} \right], \,\,\, B = \left[ \begin{array}{l}
0\\
1
\end{array} \right].
\end{align}
\noindent Let ${U_0} = \left[ {.63{\rm{ \,\,\,\,\,\,\,\, }}{\rm{.812 \,\,\,\,\,\,\,\,\,\,  - }}{\rm{.75 \,\,\,\,\,\,\,\, }}{\rm{.83 \,\,\,\,\,\,\,\, }}{\rm{.265}}} \right]$ and the initial
condition be ${x_0} = {\left[ {1{\rm{   \,\,\,\,\,\,    1}}} \right]^T}$. Then, the collected data is
\begin{align}
\nonumber {X_0} = \left[ \begin{array}{l}
1{\rm{  \,\,\,\,\,\,\,\,\,\,  - }}{\rm{.2659 \,\,\,\,\,\,\,\,\,\,   }}{\rm{.0538 \,\,\,\,\,\,\,\,\,\,  }}{\rm{.0058  \,\,\,\,\,\,\,\,\,\,  - }}{\rm{.0002}}\\
{\rm{1 \,\,\,\,\,\,\,\,\,\,  1}}{\rm{.6709  \,\,\,\,\,\,\,\,\,\, }}{\rm{.7033 \,\,\,\,\,\,\,\,\,\,  - }}{\rm{.675 \,\,\,\,\,\,\,\,\,\,    }}{\rm{.8208}}
\end{array} \right], \\ {X_1} = \left[ \begin{array}{l}
{\rm{ - }}{\rm{.2659  \,\,\,\,\,\,\,\,\,\,  }}{\rm{.0538  \,\,\,\,\,\,\,\,\,\, }}{\rm{.0058  \,\,\,\,\,\,\,\,\,\,  - }}{\rm{.0002  \,\,\,\,\,\,\,\,\,\,  0}}\\
{\rm{1}}{\rm{.6709 \,\,\,\,\,\,\,\,\,\,  }}{\rm{.7033 \,\,\,\,\,\,\,\,\,\,  - }}{\rm{.675  \,\,\,\,\,\,\,\,\,\,   }}{\rm{.8208  \,\,\,\,\,\,\,\,\,\, }}{\rm{.2675}}
\end{array} \right],
\end{align}
with the gain-scheduling data
\begin{align}
{W_0} = \left[ \begin{array}{l}
{\rm{.0488 \,\,\,\,\,\,\,\,\,\,  }}{\rm{.1392  \,\,\,\,\,\,\,\,\,\,\,  }}{\rm{.2734  \,\,\,\,\,\,\,\,\,\,\, }}{\rm{.4788  \,\,\,\,\,\,\,\,\,\,   }}{\rm{.4824}}\\
{\rm{.9512  \,\,\,\,\,\,\,\,\,\,\, }}{\rm{.8608  \,\,\,\,\,\,\,\,\,\, }}{\rm{.7266  \,\,\,\,\,\,\,\,\,\,  }}{\rm{.5212 \,\,\,\,\,\,\,\,\,\,\,    }}{\rm{.5176}}
\end{array} \right],
\end{align}
which gives
\begin{align}
{X_W} = \begin{bmatrix}
.0488 &   -.037  &  .0147 &  - .0028  &    - .0001\\
.0488  & .2327 & .1923  &   - .3232 &    .396\\
.9512  & - .2288 & .0391  &  -.003 &      -.0001\\
.9512  & 1.4382 & .511   &   - .3519 &    .4248
\end{bmatrix}.
\end{align}
The number of samples is $T=5$, which satisfies Assumption 2. However, it does not satisfy
the full row rank condition for \eqref{rank}. Let now design a safe controller for the polyhedral set in form \eqref{poly} with the matrix $F$ as \vspace{-6pt}
\begin{align} \label{F}
F=\begin{bmatrix}
1/5 & 2/5 \\
-1/5 & -2/5 \\
-3/20 &  1/5 \\
3/20 &  -1/5
\end{bmatrix}.
\end{align}
Using Theorem 6, a safe gain-scheduling controller is learned as 
\begin{align}
{K_{1,s}} = {U_0}{G_{{K_{1,s}}}} = \left[ { - .5{\rm{  \,\,\,\,\,\,\,\,\,\,  - }}{\rm{.1  \,\,\,\,\,\,\,\,\,\,   - }}{\rm{.8687 \,\,\,\,\,\,\,\,\,\,  }}{\rm{.175}}} \right],
\end{align}
with
\begin{align}
{G_{{K_{1,s}}}}=\begin{bmatrix}
-5.63 & -.0744 & 1.2847 &  .0745 \\
- 12.76 & -.5156 & .2298 &  .4952 \\
67.57 & -1.315 & -4.784 &  1.143 \\
67.88 & -1.4725 & -5.7252 &  .7227 \\
30.78 & 2.274 & -2.642 &  .2248
\end{bmatrix}.
\end{align}
Therefore, while it is impossible to learn the system dynamics using the collected data, a safe controller is directly learned.

 The following result solves Problem 1 for the case in which the control input is constrained. \vspace{3pt}

\noindent\textbf{Corollary 3.} Consider the data collected in \eqref{datU}--\eqref{datW}. Let Assumptions 1 and 2 hold. Let $u\in \cal{U}$ where
\begin{align} \label{U}
{\cal{U}} = \{ u \in {R^m}:Uu \le \bar 1\}.
\end{align}
Then, Problem 1 is solved if there exist decision variables   
${G_{{K_{1,s}}}}$ and  
${P_{1,s}} \ge 0$, such that

\begin{align} \label{d1}
\left\{ \begin{array}{l}
{P_{1,s}}({\delta _j} \otimes \bar 1) \le \lambda \bar 1,j = 1,...,s  \\
{P_{1,s}}({I_{s \times s}} \otimes F) = F{X_1}{G_{{K_{1,s}}}}\\
{X_W}{G_{{K_{1,s}}}} = I\\ 
U{U_0}{G_{{K_{1,s}}}}{D_i}k \le \bar 1{\rm{  }}, i = 1,...,s \,\, {\rm{  and }} \,\, \forall k \in {\rm{vert{\cal{S}}}(F, \bar 1)}.
\end{array} \right. 
\end{align}
Moreover,
${K_{1,s}} = {U_0}{G_{{K_{1,s}}}}$ and the control gains that solve the problem are obtained as 
${K_i} = {U_0}{G_{{K_{1,s}}}}{D_i}$. \vspace{3pt}

\noindent\textit{Proof.}  The proof is similar to Theorem 6 and is omitted. \vspace{3pt}


\noindent\textbf{Remark 6.} Note that \eqref{databasedpolyF}--\eqref{databasedpolyF2} or \eqref{d1} corresponds to solving a linear program in the decision variables  
${G_{{K_{1,s}}}}$ and
${P_{1,s}}$, hence they are numerically appealing. 
\vspace{-8pt}

\subsection{Data-Based Safe Gain-Scheduling for Ellipsoidal Sets} 
The next result provides a data-based control design procedure for LPV systems for which their safe set is described by an ellipsoidal set. \vspace{4pt}

\noindent\textbf{Theorem 7.} Consider the data collected in \eqref{datU}--\eqref{datW}. Let Assumptions 1 and 2 hold. Let
$u \in {R^m}$.  Then, Problem 2 is solved if there exist decision variables 
${G_{{K_{1,s}}}}$ such that
\begin{align} \label{elipdata}
 \left\{ \begin{array}{l}
\left[ \begin{array}{l}
{\lambda^2 }P\,\,\,\,\,\,\,\,\,\,\,\,\,\,\,\,\,\,\,\,\,\,{\rm{    (}}{{\rm{X}}_1}{G_{{K_{1,s}}}}{D_i})\\
{({{\rm{X}}_1}{G_{{K_{1,s}}}}{D_i})^T}\,\,\,\,\,{\rm{   }}P^{-1}
\end{array} \right] \succeq 0{\rm{    }}\,\,\,\,\,i = 1,...,s\\
{X_W}{G_{{K_{1,s}}}} = I.
\end{array} \right.
\end{align}

\noindent\textit{Proof.} 
The $\lambda$-contractivity condition for the LPV systems with ellipsoidal safe sets is shown in \eqref{modelelip}. Therefore, the proof is completed if one shows that the satisfaction of \eqref{elipdata} implies the satisfaction of \eqref{modelelip} with ${K_{1,s}}$. Using the Schur complement, \eqref{elipdata} is equivalent to
\begin{align}
{D_i}^T{({X_1}{G_{{K_{1,s}}}})^T}P({X_1}{G_{{K_{1,s}}}}){D_i} - {\lambda ^2}P \preceq 0{\rm{   }}\,\,\,\,\,i = 1,...,s.
\end{align}

\noindent Using the equality in \eqref{elipdata} and ${K_{1,s}} = {U_0}{G_{{K_{1,s}}}}$, \eqref{G} is obtained with ${A_{1,s}} + B{K_{1,s}} = {X_1}{G_{{K_{1,s}}}}$, based on Theorem 5. Comparing \eqref{elipdata} with \eqref{modelelip} and using ${A_{1,s}} + B{K_{1,s}} = {X_1}{G_{{K_{1,s}}}}$ completes the proof. \hfill   $\square$  \vspace{6pt}

\noindent\textbf{Remark 7.} The results of Theorems 5, 6 and 7 showed that direct leaning of a safe controller for an LPV system is highly advantages over model-based safe control deign that relies on system identification. This is highly advantageous when the LPV system has many scheduling variables and/or control inputs. \vspace{6pt}

\noindent\textbf{Remark 8.} Note that \eqref{elipdata} corresponds to solving a semi-definite program in the decision variables   %
${G_{{K_{1,s}}}}$ and   
${P_{1,s}}$, hence it is numerically appealing. Compared with polyhedral sets, however, it is computationally more demanding. \vspace{-6pt}
\section{	Numerical Example} \vspace{-3pt}

\noindent To verify our results, the following simulation example is considered.
The safe set  ${\cal{S}}(F, \bar 1)$ is defined as in \eqref{poly} with $F$ defined in \eqref{F}, and the set  $\cal{U}$  specifies the condition
 $ - 8 \le u \le 8 $. The contractivity level is chosen as  
$\lambda  = .84$. The data used for learning the controller are collected from an open-loop experiment, as shown in Figure 1, in which the control input $u$ is chosen as a random variable uniformly distributed on
$[-1,1]$. The matrices ${A_1},{A_2}$ and $B$  generating these data are \vspace{3pt}
 \begin{align} 
   {A_1} = \left[ \begin{array}{l}
1\,\,\,\,\,\,\,\,{\rm{  2/3}}\\ 
{\rm{ - 2/6\,\,\,\,\,\,\,\, 1}}
\end{array} \right],\,\, {\rm{ }}{A_2} = \left[ \begin{array}{l}
4/5\,\,\,\,\,\,\,\,\,\,\,\,{\rm{  2/5}}\\
{\rm{ - 2/5\,\,\,\,\,\,\,\,\,\  6/5}}
\end{array} \right],{\rm{ }} \,\, B = \left[ \begin{array}{l}
0\\
1
\end{array} \right].
 \end{align}

 \begin{figure}
 
    \begin{subfigure}{10 cm}
        \includegraphics[width=8cm]{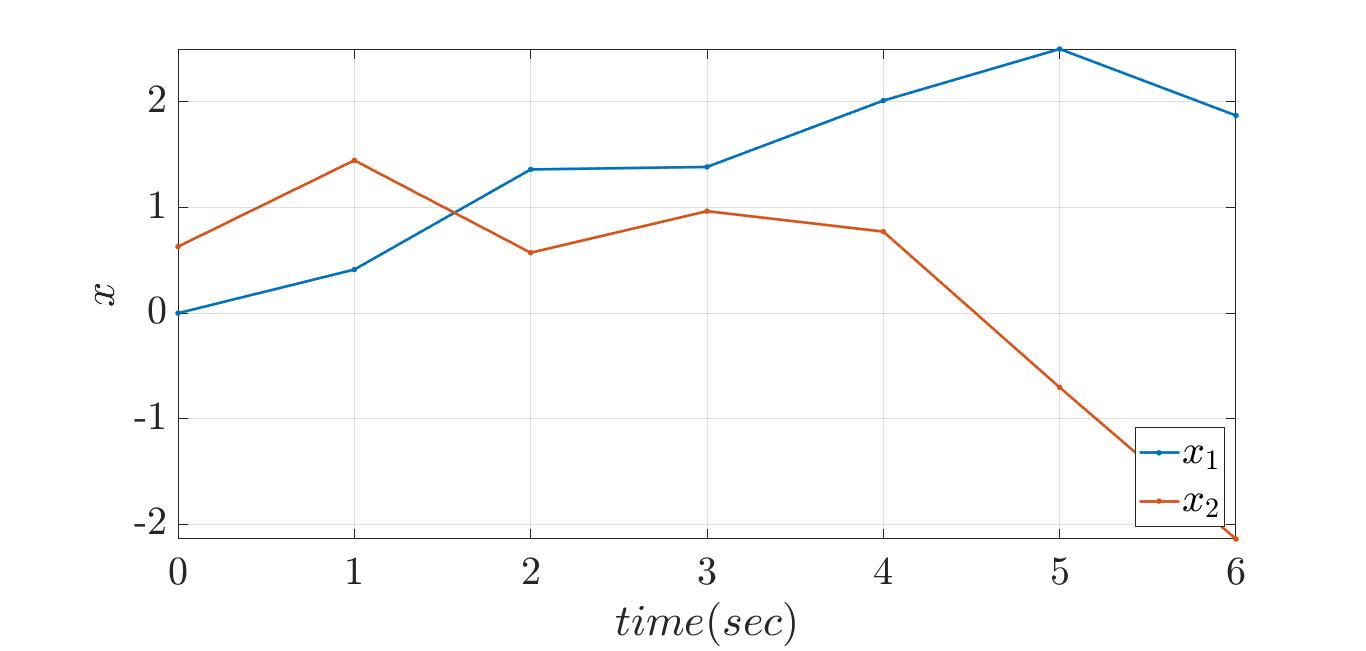}
        \caption{}
     \end{subfigure}
    \begin{subfigure} {10 cm}
        \includegraphics[width=8cm]{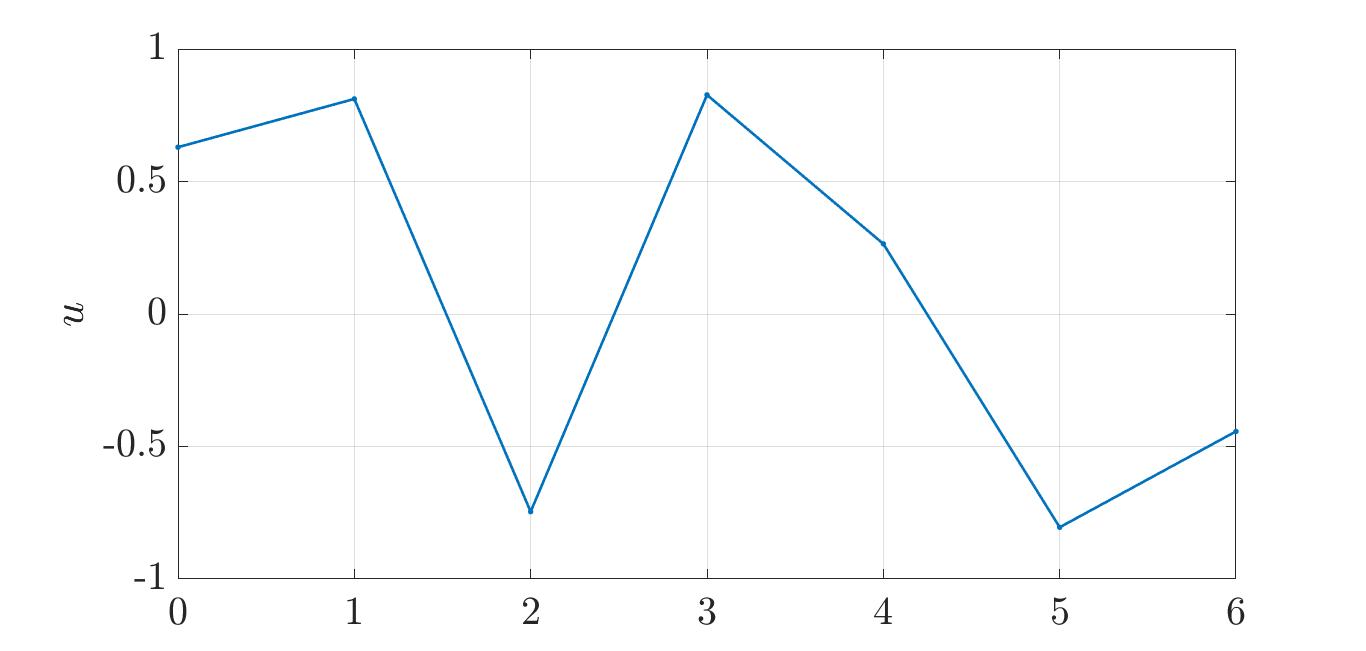}
        \caption{}
     \end{subfigure}
     \begin{subfigure} {10 cm}
        \includegraphics[width=8cm]{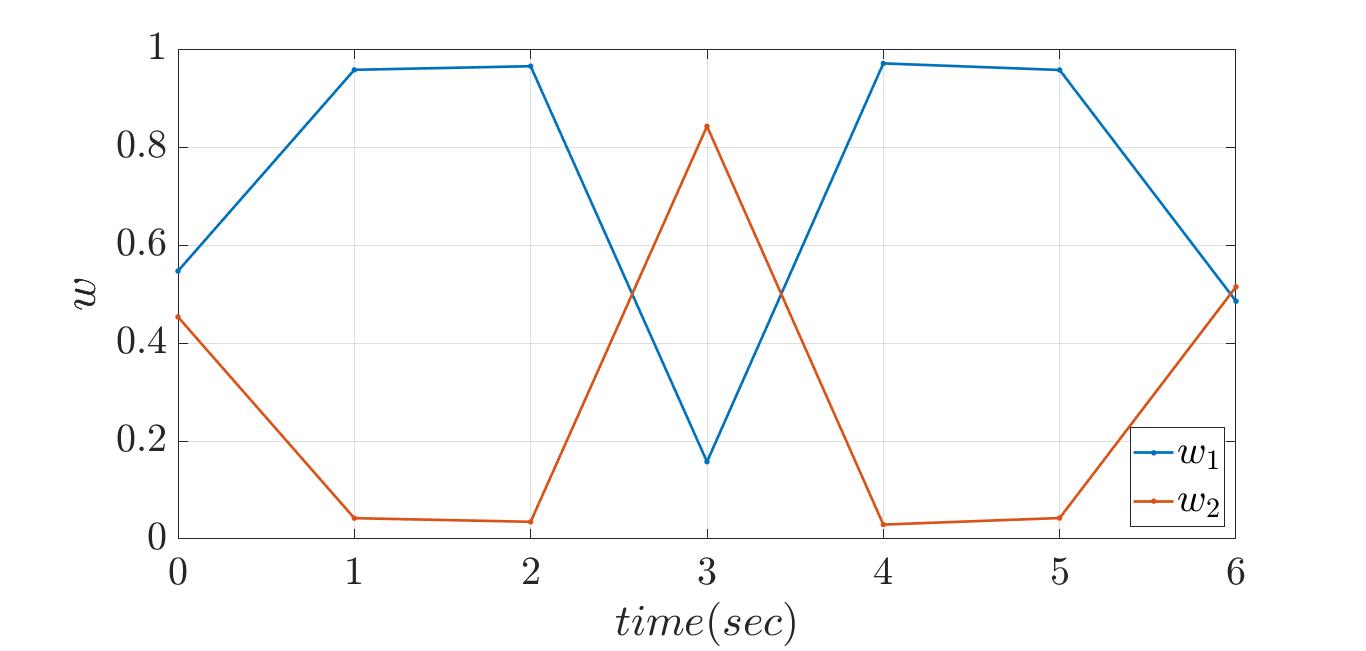}
        \caption{}
     \end{subfigure}
    \caption{(a) State sequences of data as in \eqref{DatX0} and \eqref{datX1}, with $T=6$ (b) Input sequences of data as in \eqref{datU} with $T=6$ (c) gain-scheduling sequences of data as in \eqref{datU}--\eqref{datW1} with $T=6$.}
\end{figure}
 \begin{figure}
 
    \begin{subfigure}{8 cm}
        \includegraphics[width=8cm]{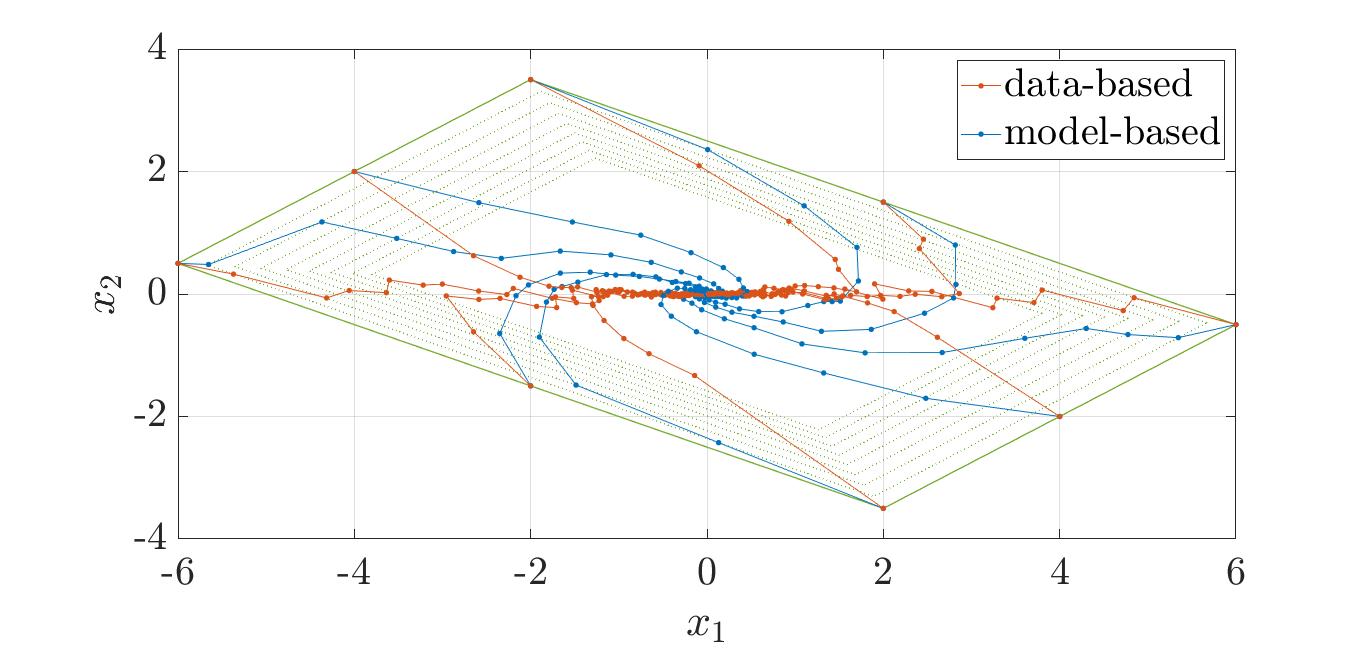}
        \caption{}
     \end{subfigure}
    \begin{subfigure} {8 cm}
        \includegraphics[width=8cm]{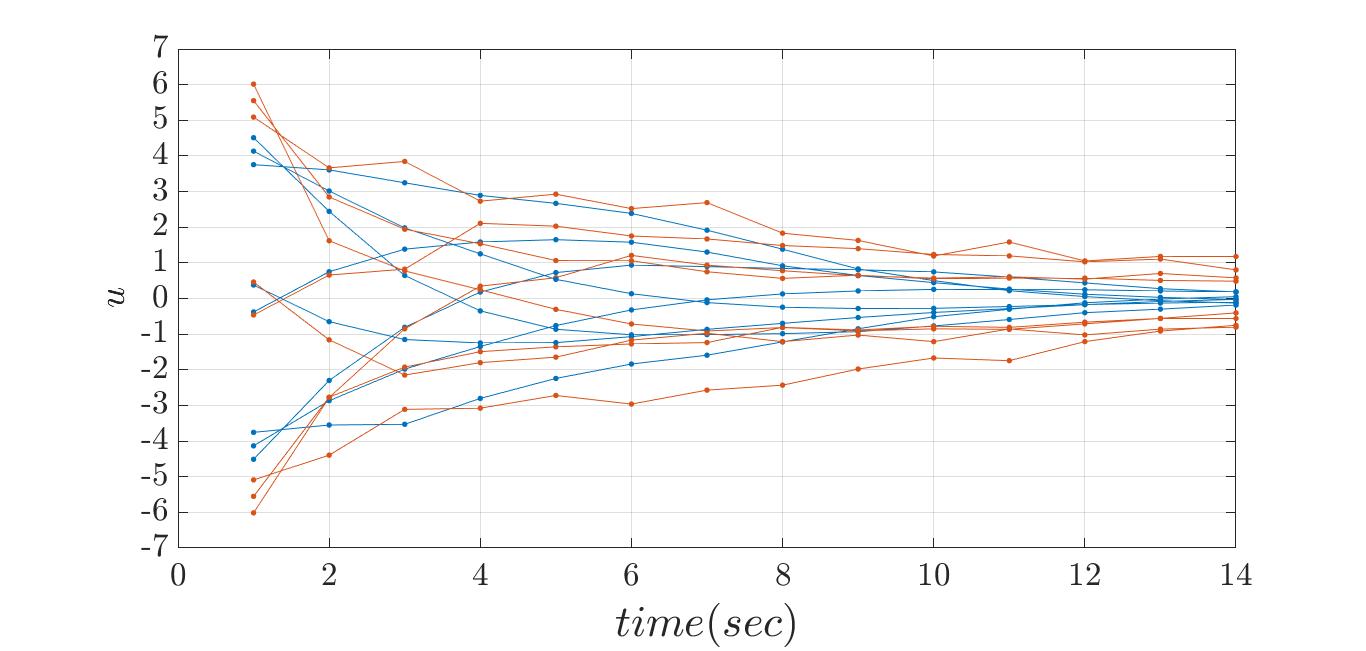}
        \caption{}
     \end{subfigure}
    \caption{(a) The safe set ${\cal{S}}(F, \bar 1)$ and solutions arising from the gain-scheduling state feedback law. (b) Control signal $u$ corresponding to the solutions in orange and blue for different initial conditions depicted in (a).}
\end{figure}

 \noindent       The linear optimization problem in Theorem 6 is solved in the variables 
${G_{{K_{1,2}}}}$ and  
${P_{1,2}}$, and the resulting  
${K_{1,2}}$ is      
         ${K_{1,2}} = \left[ {.3056,{\rm{   - }}{\rm{.3889  }},{\rm{.2389,   - }}{\rm{.5889}}} \right]$. Only for illustrative purposes, we also solve the model-based safe control design conditions \eqref{modelpoly} and    
       obtain a gain matrix
  ${K_{1,{2_{A,B}}}} = \left[ {.2680,{\rm{-}}{\rm{.8398,  }}{\rm{.4722,-}}{\rm{.4556}}} \right]$. 
The safe set is shown with a green solid line in Figure 2. Figure 2a shows the states of the system for different initial conditions resulting from both the data-based controller (orange) and the model-based controller based on the classical model-based approach (blue). The set $\cal{S}$ (green, solid) and the sets ${\lambda \cal{S}},{\lambda ^2}{\cal{S}},{\lambda ^3}{\cal{S}},...$ (green, dotted) are also shown. This shows that safety is guaranteed as the states only evolve in the safe set. Figure 2b also certifies that 
 the control signal satisfies the constraints given by $\cal{U}$. \vspace{-6pt}

\section{conclusion} \vspace{-2pt}
\noindent This paper presents a data-based solution to the safe gain-scheduling control problem. The presented solution finds a safe controller for nonlinear systems represented in LPV form while only relying on measured data, and it is shown that it enforces not only stability but also invariance with a given polyhedral or ellipsoidal set. The presented data-based solution results in numerically eﬃcient linear program for polyhedral sets and semi-definite program for ellipsoidal sets. \vspace{-10pt}

\bibliographystyle{IEEEtran}
  \vspace{-6pt}

\end{document}